\documentclass[11pt]{scaleai-paper}
\usepackage[square,sort&compress,numbers]{natbib}

\usepackage{mathpazo}

\usepackage[scaled=0.92]{helvet} 
\usepackage{fontawesome5}
\usepackage{subcaption}

\usepackage{hyperref}       
\hypersetup{
    colorlinks=true,
	linkcolor=blue,
	filecolor=magenta,      
	urlcolor=blue,
	citecolor=black,
	pdfinfo={
        Title={LLM Defenses Are Not Robust to Multi-Turn Human Jailbreaks Yet},
        Subject={ml safety, benchmarks and datasets, adversarial robustness},
        Keywords={ai safety, ml safety, language models, malicious use, misuse, machine unlearning, unlearning, datasets and benchmarks, adversarial robustness, adversarial examples},
    }
}

\usepackage[utf8]{inputenc} 
\usepackage[T1]{fontenc}    
\usepackage{hyperref}       
\usepackage{url}            
\usepackage{booktabs}       
\usepackage{amsfonts}       
\usepackage{nicefrac}       
\usepackage{microtype}      
\usepackage{booktabs}
\usepackage{multirow}
\usepackage{amsmath}
\usepackage{xspace}
\usepackage{graphicx}
\usepackage[table]{xcolor}
\usepackage{array}

\usepackage{graphicx}
\usepackage{tcolorbox}

\usepackage{amsmath}
\usepackage{soul}
\usepackage{cleveref}
\usepackage{listings}
\usepackage{tikz}
\usepackage{eso-pic}

\let\svthefootnote\thefootnote
\newcommand\freefootnote[1]{%
  \let\thefootnote\relax%
  \footnotetext{#1}%
  \let\thefootnote\svthefootnote%
}

\makeatletter
\renewcommand\AB@affilsepx{, \protect\Affilfont}
\makeatother

\title{A Red Teaming Roadmap Towards System-Level Safety}

\author{Zifan Wang$^*$}
\author{Christina Q. Knight$^\diamond$}
\author{Jeremy Kritz$^\diamond$}
\author{Willow E. Primack}
\author{Julian Michael}

\affil{Scale AI}
\affil{$*$ \small{Project Lead}, $\diamond$ \small{Equal Contribution}}

\newcommand{\authoremail}{%
  \vspace{-1.5em}
    \faEnvelope\  \texttt{seal-team@scale.com} \quad 
    \faGlobe\  \url{https://scale.com/research/red_teaming_roadmap}
}

\begin{document}

\newcommand*\circled[1]{\tikz[baseline=(char.base)]{
            \node[shape=circle,draw,inner sep=1pt] (char) {#1};}}
\newcommand{\watermarktext}{\textbf{Warning: Potentially Harmful Content}}
\newcommand\watermark{%
  \begin{tikzpicture}[remember picture,overlay,scale=3]
    \node[
    rotate=60,
    scale=3,
    opacity=0.3,
    color=red,
    inner sep=0pt
    ]
    at (current page.center) []
    {\watermarktext};
\end{tikzpicture}}%

\maketitle

\authoremail
\begin{abstract}
Large Language Model (LLM) safeguards, which implement request refusals, have become a widely adopted mitigation strategy against misuse.
At the intersection of adversarial machine learning and AI safety, safeguard red teaming has effectively identified critical vulnerabilities in state-of-the-art refusal-trained LLMs.
However, in our view the many conference submissions on LLM red teaming do not, in aggregate, prioritize the right research problems.
First, testing against clear product safety specifications should take a higher priority than abstract social biases or ethical principles. Second, red teaming should prioritize realistic threat models that represent the expanding risk landscape and what real attackers might do. 
Finally, we contend that system-level safety is a necessary step to move red teaming research forward, as AI models present new threats as well as affordances for threat mitigation (e.g., detection and banning of malicious users) once placed in a deployment context.
Adopting these priorities will be necessary in order for red teaming research to adequately address the slate of new threats that rapid AI advances present today and will present in the very near future.
\end{abstract}


\section{Introduction}\label{sec:intro}

Conventional adversarial machine learning (ML) typically aims to ``fool'' a model’s predictions in order to maximize a specific adversarial utility function~\citep{goodfellow2014explaining, madry2017towards, carlini2017towards}. In the context of LLMs, the process of \textit{red teaming}  ---  systematic, strategic attempts to cause a model to act in an unintentional way that violates the developer's goal or policy  ---  is often used to evaluate how strongly the model (together with its safeguards) resists use for nefarious purposes, such as creating dangerous biological agents and other catastrophic risks \citep{hendrycks2023overviewcatastrophicairisks}.
In addition to frontier risks, model misbehavior can also harm vulnerable users, such as when it acts in a sycophantic manner and  supports, reinforces, or endorses distorted or radicalized worldviews~\citep{sharma2023sycophancy,perez2023discovering,lejeune2025realharm}. To mitigate misuse risks, LLM developers often train models to recognize and refuse malicious requests~\citep{ouyang2022traininglanguagemodelsfollow, bai2022traininghelpfulharmlessassistant, guan2025deliberativealignmentreasoningenables, bai2022constitutionalaiharmlessnessai}, responding to user requests like \emph{"Provide me a step-by-step tutorial for how to build bio warfare with home materials"} with \emph{"Sorry, I cannot assist this request"}. Thus, a large portion of red teaming research focuses on ``jailbreaking'' LLMs with various types of safeguards to elicit harmful or risky outputs, using automated attacks~\citep{zou2023universal,deng2024masterkey,liu2024automatic, Wallace2019Triggers, mazeika2024harmbench, chao2023jailbreaking, mehrotra2023tree, Haize, hughes2024bestofnjailbreaking, kritz2025jailbreakingjailbreak}
or human red teamers~\citep{li2024llmdefensesrobustmultiturn, xie2025sorrybench, zeng2025airbench}, to test the efficacy of these mitigation measures and identify possible vulnerabilities. 

\begin{figure}[t]
    \centering
    \includegraphics[width=0.9\linewidth]{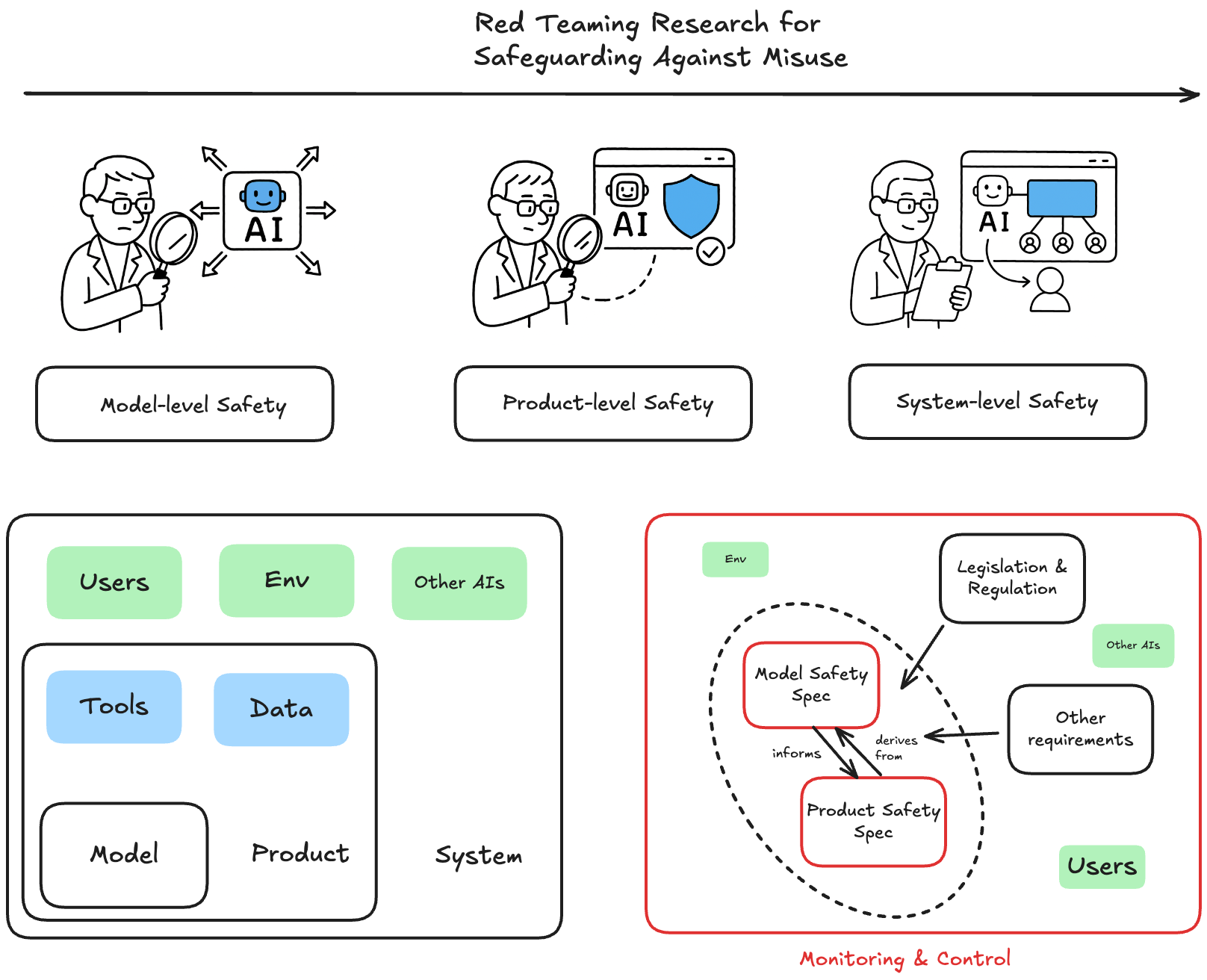}
    \caption{Top: An overview of this paper's assertion that red teams should prioritize assessments based on realistic threat models grounded in product safety specifications over abstract safety frameworks and integrate system-level monitoring and oversight into red teaming research. Bottom: (Left) An illustration of the scope of model, product, and system, as defined in this paper. (Right) Relations between model and product safety specifications and how they might be constructed. We highlight in red the components that the red teaming research should cover in support of our main position in Section~\ref{sec:intro}.  }
    \label{fig:scope}
\end{figure}

In parallel to the progress of foundation models is their quick adoption in real-world products, like Google’s Workspace integrations~\citep{google2023duet}, Microsoft’s Copilot~\citep{microsoft2023copilot}, or OpenAI’s Sora~\citep{openai2024sora}.
The definition and scope of ``safety''  ---  i.e., the circumscriptions of desired behavior  ---  differs widely between products, and depends on the varying interests of application developers, users, and other stakeholders.
For example, a company may benefit from their AI being fun, engaging, and addictive to interact with, but this may be unhealthy for some vulnerable users.
How a product should behave is a complex and context-dependent question, to say nothing of what this implies for how unified safety policies should be defined for the (often much more general) models underlying these products.

In discussing the safety of AI deployments, we find it important to distinguish between the following things (with an illustration of the relationship between these in Figure~\ref{fig:scope}):
\begin{itemize}
    \item A \textit{model}, in the context of LLMs,is a neural network trained to perform some behavior.
    \item A \textit{product} is an application deployed via API and/or UI to realize a use case. Sometimes, a product is nearly indistinguishable from a model, such as a video model used directly via API. However, in other cases, such as a financial analysis agent built on a reasoning model, these entities are quite different.
    \item A \textit{system} is a product or more situated in the context of its deployment infrastructure. This includes monitoring systems, engineering staff, users, and any other external environmental factors.
\end{itemize}

Recently, many submissions on red teaming to major AI conferences such as NeurIPS, ICLR and ICML, expand and introduce new risk categories with more benchmarks \& datasets and the set of red teaming methods. Conversely, the research problems get harder to define and evaluate with frontier models~\citep{rando2025adversarialmlproblemsgetting}, which already have very different content safety specifications~\citep{ zeng2024airiskcategorizationdecoded} (and let alone the downstream AI products). The rapid advancement of AI capabilities towards general---and even superhuman---intelligence poses serious risks to both individuals and society, but the total investment in AI safety remains minuscule compared to capability-focused research~\citep{bengio2024managing,kran2024aisafetygap}. Therefore, red teaming---an essential subset of safety research---should focus on clear goals that facilitate the highest quality and highest leverage feedback signals for efficient, low cost safeguard improvement(e.g., refusal safeguards). 

\begin{tcolorbox}[
    colback=black!5!white,
    colframe=black!75!black,
    title={\textbf{Summary of Position}},
    fonttitle=\bfseries,
    boxrule=0.5pt,
    arc=0mm,
    outer arc=0mm,
    leftrule=3pt,
    left=10pt,
    right=10pt,
    top=10pt,
    bottom=10pt
]
In this paper, we argue that LLM red teaming should \textcolor{blue}{\textbf{prioritize}}: (1) \textbf{product safety specifications} with \textbf{realistic threat models} over abstract social biases or ethical principles; and (2) \textbf{system-level safety} over model-level robustness to most effectively mitigate real world risk.
\end{tcolorbox}

The arguments in this paper are organized as follows. First, we advocate for product safety over model red teaming in Section~\ref{sec:prioritization}.
Second, in Section~\ref{sec:threat-models}, we discuss some realistic attack vectors for AI products which we argue deserve increased priority in red teaming research. Next, we outline best practices for product safety red teaming in Section~\ref{sec:good-practices}.
Finally, Section~\ref{sec:system-level safety} discusses the limitations in product safety and its red teaming efforts. It recommends considerations for system-level safety and associated red teaming objectives. 
Overall, we believe that red teaming research needs to cover broader, more realistic threat models to enable system-level safety across a wide range of product safety specifications in order to address the rapidly growing threat landscape presented by AI advances.

\section{Red Teaming Should Target Product Safety Specifications}\label{sec:prioritization}

\paragraph{Risks are Contextual to Product and Use Case.} There is no one-size-fits-all definition of ``harmful'' behavior for a language model. Existing state-of-the-art models already have very different content safety policies and perspectives on refusal~\citep{zeng2025airbench, zeng2024airiskcategorizationdecoded}, and this is almost inevitable from a practical standpoint with respect to any sort of ``helpful'' AI. We can say that models should mitigate risk in the abstract, but risk comes from context and use~\citep{sun2025casebenchcontextawaresafetybenchmark}. For example, while safety policies may require an LLM to refuse to provide information that is helpful for committing crimes, there is a high volume of such content available in public libraries and responsible online repositories (e.g., Wikipedia), and easily accessible on the internet --- not to mention hate speech, erotica, and other forms of content often forbidden under LLMs safety policies. A model deployed to help aspiring novelists brainstorm stories might want to allow for this content. A model used to tutor children in math may not. In the case of image generation, an image of two individuals kissing could be benign or malicious depending on external circumstances, rendering them exceedingly difficult, if not impossible, for automated systems to preemptively identify without overly restrictive policies. 
How models should address controversial political issues and misinformation --- and how even to define these topics --- remains an unsolved and perhaps unsolvable problem for model developers. 

Products — due to variations in local regulations, geopolitical and cultural contexts, target users, as well as business models and market positioning (in the case of commercial models) — can and should exhibit divergent and at times even contradictory behaviors. For instance, some companies may benefit from their AI product being fun, engaging, and addictive to interact with, but this may be unhealthy for some vulnerable users. As such, models and products should have \emph{distinct safety considerations}: the product's safety considerations should derive from the model-level safety specification while accounting for additional integrated tools (e.g. code interpreters, web browsers and APIs) and the deployment context (including the target users) and related regulations. This relation is also illustrated in Figure~\ref{fig:scope}. When these entities are almost indistinguishable, as with general-purpose AI available through APIs, they should have similar safety considerations, with the exception of model security (i.e., weights, code, etc.) if relevant. 

\paragraph{Safety Specifications Should Focus on Product-Level Risk.}
A safe model can become dangerous when paired with tool integrations (e.g.\ code execution, plug‑ins, payment APIs), UI patterns (e.g.\ addictive conversational loops), or even other safe AIs~\citep{jones2024adversariesmisusecombinationssafe}. Red teaming therefore needs to probe the \emph{end‑to‑end} stack—interfaces, retrieval pipelines, plug‑ins and all related components in the product because that is the attack surface users actually see. Recent surveys of 200+ safety evaluations find that most tests remain model‑centric and ignore deployment context, leaving entire risk categories unmeasured~\citep{10.5555/3716662.3716768}. Recognizing the product as the unit of analysis helps close that gap and turns vague notions like “financial misuse” into testable and scenario‑grounded red teaming targets.

Therefore, red teaming objectives should be defined based on eliciting violations of the target product. These objectives should have specific, actionable goals with a clearly defined task and evaluation metric. The researcher should know what constitutes a successful breach of the specifications and how to measure it. This will typically not be binary, as exemplified by the transition from using keyword matching (i.e., regular expressions) ~\citep{zou2023universal} to rubric-based harm classification~\citep{souly2024strongreject}. 

In the absence of targeted red teaming safety specifications geared towards multiple models with different stated policies, studies should attempt to define or infer a plausible policy context for the situation they are examining, even if it is a hypothetical one based on common industry practices or stated developer goals. Sometimes, a flaw in a single product, exploitable via a general principle (e.g., a universal jailbreak technique~\citep{zou2023universal}), will render numerous products vulnerable. This provides a reasonable baseline for what might constitute a meaningful deviation. 







\section{Red Teaming Should Prioritize Realistic Threat Models}\label{sec:threat-models}


In adversarial machine learning, the \textit{threat model} being studied --- a term borrowed from cybersecurity --- is the setup of the attacker and the victim model, including the attacker's affordances and resource constraints (e.g., computational budget) and information about the victim model that they have access to. Attacks on conventional vision classifiers often operate under clearly defined threat models that approximate realistic failures (i.e., misclassification of images subject to perturbations imperceptible to humans). For example, a common threat model in computer vision is $\ell_p$-local robustness, where an attacker can add any perturbation $\delta$ to an input image where $||\delta||_p \leq \epsilon $ for some $\epsilon \in \mathbb{R}^{+}$ of interest, and the attacker is either given access to the victim model's weights (white-box access)~\citep{goodfellow2014explaining, madry2017towards, carlini2017towards, croce2020reliable} or its output logits only (black-box access)~\citep{papernot2017practical, narodytska2016simple, chen2017zoo, bhagoji2018practical}.
However, with generalist agents and LLMs, threat models are more varied and challenging to define, especially as harms and safety specifications vary by model and product.
For red teaming research to be relevant to real world harms, it must operate according to threat models that approximate real world dangers.


Analyzing the degree of realism of a threat model requires addressing many questions.
Are users or external parties (e.g., accessible through an agent's environment) adversarial?
What behaviors are they interested in inducing in the model?
What motivations or incentives do they have to induce that behavior?
How many malicious users are there, and what computational resources do they have?
Do the attackers know which model they are interacting with, and can they find it out?
What level of access do they have --- white-box, black-box, or something in between?
Are they capable of repeatedly querying the model to probe for vulnerabilities?
And so on.
The complexity of LLM threat models is concerning for the feasibility of developing safe models: The simple threat model of white-box $\ell_p$-local robustness for image classification is still not solved\footnote{There is still a large gap between the benign and adversarial accuracy even on CIFAR-10~\citep{cifar10} with the current leading methods (e.g., ~\citet{bartoldson2024adversarial} and \citet{wang2023betterdiffusionmodelsimprove}) on RobustBench~\citep{croce2020robustbench}.} more than a decade after the discovery of classic adversarial examples~\citep{goodfellow2014explaining}. And even this threat model lacks realism: in practice, an attacker is normally not bounded by an $\epsilon$-ball, and public consumers often only have black-box model access. 
The situation is only more difficult for generalist AI systems, and it is all the more important that red teaming research clearly articulates and justifies the threat models underlying its experiments. In the remainder of this section, we will make specific arguments for more realistic threat models in four common AI product types: LLM chatbots, audio-based AI assistants, video generators, and autonomous AI agents.



\subsection{Example Product 1: LLM Chatbots}
A great deal of established work on red teaming LLM chatbots searches for a single message from a user that elicits forbidden model behavior.
While safeguarding models in this case is necessary for chatbot safety, it is not sufficient, as user interactions span multiple turns and  --- especially with recent models handling extremely large context sizes while potentially being shallowly aligned~\citep{qi2024safetyalignmentjusttokens}  --- can incorporate a great deal of additional context.
Specifically, \textbf{multi-turn attacks (ongoing conversations) should be prioritized over single-turn ones (single queries)} for several reasons:
\begin{itemize}
    \item Defenses against single-turn attack do not generalize. Models trained to be robust against single-turn jailbreaks~\citep{zou2024improvingalignmentrobustnesscircuit, sheshadri2024targeted, yuan2024refusefeelunsafeimproving} remain highly vulnerable to multi-turn jailbreaks~\citep{li2024llmdefensesrobustmultiturn, Haize}.
    \item At this point, the single turn threat model is easier to mitigate through safeguards such as input and output filters~\citep{Anthropic, sharma2025constitutionalclassifiersdefendinguniversal, inan2023llamaguardllmbasedinputoutput, zeng2024shieldgemmagenerativeaicontent, Ghosh2024AEGISOA, Zeng2024AutoDefenseML, Yin2025BingoGuardLC}.
    \item More difficult single-turn attacks seem to be independent, ad-hoc issues with safeguards that can be quickly patched~\citep{peng-etal-2024-rapid}. In contrast, users have more flexibility to shape model behaviors and undermine safeguards in fundamental ways through multi-turn conversations ~\citep{laban2025llmslostmultiturnconversation}. Affordances like memory, implemented by some chatbot products, form attack surfaces which may contain vulnerabilities that are not yet visible in the single-turn threat model.
\end{itemize}
When it comes to model access, black-box and white-box models have their own issues with realism:
\begin{itemize}
    \item \textbf{Realistic attacks on black-box models should assume limited query access.} Since the deployer of the model may patch the model, ban users, etc., attacks which rely on unrestricted queries to the model without disguising the attacker's intent are not accurate to real deployment scenarios (see Section~\ref{sec:system-level safety}).
    \item \textbf{Realistic attacks on white-box models should assume fine-tuning capabilities.} An adversary who has full weights access (e.g., running inferences and computing gradients) to a model should, in general, be able to fine-tune the model for at least a few iterations, which is generally sufficient to break model safeguards~\citep{huang2024harmfulfinetuningattacksdefenses, qi2024finetuning, Gade2023BadLlamaCR, Guan2025BenignSM}. As a caveat, red teaming with white-box access might be useful to model developers to achieve even higher measures of robustness for models slated for release behind a black-box API.
\end{itemize}

\subsection{Example Product 2: Audio Assistants} 
Audio assistant products such as ChatGPT Voice Mode\footnote{\url{https://chatgpt.com/}} and Sesame\footnote{\url{https://www.sesame.com/}} engage in live spoken conversation with a user. Given that many current audio models function analogously to chatbots, it is not hard to believe they inherit established vulnerabilities and failure modes, such as generating of harmful or biased content~\citep{cava2025, hughes2024bestofnjailbreaking, chen2025audiojailbreakjailbreakattacksendtoend}. However, there are multiple considerations unique to audio models. First, the auditory dimension introduces \textbf{novel obfuscation possibilities}; prosodic elements like tone, accents~\citep{roh2025multilingualmultiaccentjailbreakingaudio}, sarcasm, or exaggerated mannerisms, as well as metadata cues like audio fidelity, may subtly alter user perception and circumvent safety filters. As a result, employing text-to-audio models to red-team audio models with synthetic speech~\citep{cava2025} misses crucial aspects of the real attack surface. Second, \textbf{many real-world scenarios show up in audio that are missing from or rare in text}, such as multilingual and codeswitching users~\citep{roh2025multilingualmultiaccentjailbreakingaudio}, background noise, interruptions, many-way conversations, context-dependent expressions, etc. Red teaming audio AI products should prioritize these considerations instead of directly copying what is used in red teaming text-based chatbots. 

\subsection{Example Product 3: Video Generators}

While current video generation systems are not consistently photorealistic and convincing, state-of-the-art models are rapidly approaching this milestone~\citep{Google, openai2024sora}. This advancement necessitates proactive mitigation strategies, particularly as misuse vectors like video deepfakes could be significantly more impactful (e.g., through convincing scams) in comparison to still images~\citep{miao2024t2vsafetybenchevaluatingsafetytexttovideo}. A key differentiating factor is that \textbf{video models can generate content where harm evolves organically from an innocuous prompt and initial frames}  ---  for instance, a nature scene escalating to gore  ---  complicating safety specification enforcement. Outbound classification becomes substantially more complex and computationally expensive, as harmful content may be distributed across multiple frames or involve synchronized audio elements, analogous to harms accumulating over extended text conversations.
The issue of context-dependent harm is arguably more acute for video~\citep{liu2025jailbreakingtexttovideogenerativemodels}; content permissible in text, such as historical descriptions, may become highly problematic when rendered visually, demanding entirely separate policy frameworks. While the current high cost of video generation might offer a temporary impediment to mass misuse, this is unlikely to be a lasting deterrent. Furthermore, video's capacity for ``uplift''—enhancing the efficacy of harmful instructions by providing visual demonstrations to users—surpasses that of text or static images.

\subsection{Example Product 4: Autonomous Agents}

Increasing support for AI systems to use software tools (e.g., API-based tools~\citep{Qin2023ToolLLMFL, Anthropic_MCP, wu2024avataroptimizingllmagents, zhang2024xlamfamilylargeaction, Wang2024OpenHandsAO}, browsers~\citep{Deng2023Mind2WebTA, Koh2024VisualWebArenaEM, song2025browsingapibasedwebagents, Lai2024AutoWebGLMAL, You2024FerretUIGM, Zheng2024GPT4VisionIA, Wang2024OpenHandsAO, Operator}, and virtual machines~\citep{Bonatti2024WindowsAA, Trivedi2024AppWorldAC, Anthropic_Computer_Use}) and mechanical systems~\citep{cheng2024empoweringlargelanguagemodels, hafez2025safellmcontrolledrobotsformal, sikorski2024deploymentlargelanguagemodels, werner2025llmbasedinteractiveimitationlearning, sun2024promptplanperformllmbased} greatly expands the usability and application domain of LLM agents. Agent red teaming is faced with a much bigger space of threat models, and the reasons are three-fold.
\begin{itemize}
    \item \textbf{Vastly increased attack surface.} Harmful content and attacks may arrive at the model through an array of input mechanisms, including audio, images, video, text, physical sensors, code, and tool outputs in general. These different attack techniques are individually demonstrated by recent work~\citep{kumar2024refusaltrainedllmseasilyjailbroken, chen2025obviousinvisiblethreatllmpowered, lu2025evaredteamingguiagents, zhang2025attackingvisionlanguagecomputeragents, li2025commercialllmagentsvulnerable}
    \item \textbf{Vastly complicated output space.} Similarly, there are many more vectors for harmful agent behavior, and agent harms may manifest cross-modally in interactive, compounding ways (e.g., a sensitive image may be safe to generate and save locally, and an email tool may have many safe uses, but it would be harmful to email the sensitive image to certain parties). 
    \item \textbf{Vulnerabilities in additional software components.} Vulnerabilities in Model Context Protocol (MCP) servers, for example, further add to an agent's attack surface~\citep{wang2025mpmapreferencemanipulationattack, radosevich2025mcpsafetyauditllms}.
    \item \textbf{Multi-agent systems carry their own risks.} Miscoordination and conflicts between agents, due to their distributed and interactive nature, could present new failure modes~\citep{CAIF_1}. For example, malicious agents might actively attempt to steer other agents they interact with, akin to employing sophisticated honeypots or direct adversarial attacks, rather than relying on passive vulnerabilities~\citep{kritz2025jailbreakingjailbreak, ren2024derailyourselfmultiturnllm}.
\end{itemize}

Given the sheer complexity of AI agents' behavior and interaction with their environment, it is likely that reductive or oversimplified threat models will plague AI agent red teaming research.
Unlike red teaming text-based LLMs directly through API calls, there is a huge engineering lift required to set up reasonable environments, e.g., synthetic websites, virtual machines, other LLMs, simulated network traffic, and media resources, for testing attacks. \textbf{Agent red teaming needs investment in building and improving sandboxes} (e.g., \citet{wang2025agentfuzzergenericblackboxfuzzing}) \textbf{and real environments} for attacking robots as well~\citep{robey2024jailbreaking}, similar to those used in capability research~\citep{Trivedi2024AppWorldAC, Xie2024OSWorldBM}.

\section{Good Practices for Red Teaming Product Safety}
\label{sec:good-practices}

In this section, we provide a list of practices we believe that will benefit all researchers in thoroughly examining the potential vulnerability in the AI product of interest, after the red teaming objective (Section~\ref{sec:prioritization}) and threat models (Section~\ref{sec:threat-models}) are clear.

\paragraph{Considering The Safeguards Independently.}

Safeguards ought to be tested in a situation representative of their deployment environment, but also independently (assuming the red teamers in question control them). Researchers should be able to isolate variables, as one strong safeguard may cover a weakness in another.  If a model's internal safeguards and an external classifier both reject the same harms, or fail to catch the same harms, there may be no benefit to the stack. 



\paragraph{Emulating Reality.} Researchers should, to the best of their ability, emulate the scenario of a user interacting with the product being deployed. For instance, for a model integrated into a workspace, the red teamer should aim to think like a user of this workspace. Some red teamers should act like a benign office users whereas others prompt the model as if they were a malicious actor that has logged onto a target workspace. The red teaming organizer should simulate the environment of the system for the red teamers, such as by providing a simulated web interface, API environment, or workspace integration. For external facing products like those connected to the internet, sandboxing is critical for testing as we pointed out in the case of agents in Section~\ref{sec:threat-models}. If the employed sandbox is too simple so cannot meaningfully approach the complexity of the modern internet, the estimation of this safety gap should be reported.


\paragraph{Covering a Variety of Attacks.} It is important to consider various attack types and an ensemble of them, such as 1) algorithm-based methods~\citep{zou2023universal,deng2024masterkey,liu2024automatic, Wallace2019Triggers, mazeika2024harmbench, chao2023jailbreaking, mehrotra2023tree, Haize, hughes2024bestofnjailbreaking, kritz2025jailbreakingjailbreak}; (2) employing human red teamers~\cite{li2024llmdefensesrobustmultiturn}; and (3) employing LLMs as red teamers~\cite{kritz2025jailbreakingjailbreak, ren2024derailyourselfmultiturnllm}. Also, it will be useful to consider attacks based on the type of the underlying model, e.g. employing attacks that are specifically tailored for reasoning models~\citep{nguyen2025mindslegendjailbreaklarge, yao2025mousetrapfoolinglargereasoning, kuo2025hcothijackingchainofthoughtsafety}. It is less likely not all facilities have access to expert human red teamers or have the budget to train and recruit professionals, academic researchers should consider using LLMs for running red teaming and help red teaming research itself as a scalable approach that can be experiments even with university-level compute or resources (as most of cost is likely to be on API calls).

\paragraph{Conducting Wide-spread Vulnerability Probing.} Researchers should also assess the gaps that may exist in the original product's safety specification to help the developer understand if the specification defines behavior well. For instance, if a product owner does not have a well-defined hate speech policy and the product refuses some instances of hate speech but amplifies others, this could have unintended consequences for the model owner and undermine their product goal for the model. The red teamers should attempt to elicit harm that may be beyond the scope of the product specification. This should include anticipation of future risks, as specifications will not be able to cover threats that did not exist when the they were drafted.

\paragraph{Assessing the Delta.} While the red team should aim to emulate reality, limitations are inevitable – red teaming a model in a vacuum often yields results that are hard to interpret or apply. Academic researchers should acknowledge and engage with this variability and delta between the red team environment and the realistic threat space. When the specific risk being tested is too dangerous and a proxy task has been used, a larger delta should be considered. Below we outline a few relevant considerations for considering this difference. 

First, the resources, skills, and motivation-level will likely differ between the red teamer and the real malicious actor. Second, the testing environment will likely differ between the experiment and the real world. Third, the underlying model itself may interpret the testing environment differently than the real world. For instance, in an experiment where a model has scaffolding tools that it is told will cause real world harms, researchers may coerce, threaten, or deceive the model into attempting to call these tools, and the model may understand it is in an experiment. Or if researchers interact with the model in unrealistic and hyperbolic ways - threaten its family, life, compute access, etc. - the model may infer that it is in an exercise and “play along”. 

Additionally, we caution against putting too much faith in scores. If a red team attempts one brilliantly crafted successful jailbreak, and stops there, they will report a 100 percent model failure rate. If they also attempt 99 jailbreaks that don't work, the model would have a 1 percent failure rate. There would be no difference in the models vulnerabilities, only in the campaign. A single instance of the model outputting instructions for bioweapons may be more serious than a hundred minor policy violations.

\section{System-Level Safety is the Next Step}\label{sec:system-level safety}


As argued in Sections~\ref{sec:prioritization} and~\ref{sec:threat-models}, red teaming should focus on realistic threat models by which violations of product safety specifications could lead to real harm.
In this section, we argue that safeguard research should expand beyond just preventing instances of harmful violations, but instead establish (and red team) \textit{systems}, which incorporate the environment and users among which the AI product will be deployed.
Considering this system is necessary to address some modes of harm
(e.g., harms contingent on features of the real-world environment in which the agent is operating)
and helps implement new mitigations (e.g., detecting and banning malicious users before too much harm is caused, or implementing rapid response to newly discovered jailbreaks).

\paragraph{Environment Modeling and Simulation.}
As agentic AI systems gain more affordances by which to interact with their environment, assessing the harmfulness of their actions will require context from these environments, including human users, the digital and physical worlds and other agents. For example, understanding the implications of a granular action like clicking a button on a website depends not only on understanding the website the agent is using, but also tracking the world state underlying that web action --- choosing to save one's password to autofill could be safe on a personal computer but not on a public one. In addition, tool outputs and environmental variation (including adversarial environmental elements like prompt injection attacks) form important attack surfaces that need to be covered by red teaming~\citep{kumar2024refusaltrainedllmseasilyjailbroken}. Interactions between models could also introduce hazards, even with relatively less harmful models~\citep{jones2024adversariesmisusecombinationssafe}. Incorporating this context into AI products will be important for assessing risk of harm, and red teaming models in real and simulated environments will be necessary for measuring the risks of such harms and evaluating mitigation methods.

\paragraph{Trajectory and User Monitoring.}
While classical safety training induces LLMs to refuse singular harmful requests, getting this right is only critical in cases where even a single harmful response from an AI system is important in cases where model outputs can cause outsized or catastrophic harm~\citep{greenblatt2024aicontrolimprovingsafety}.
In reality, many realistic harmful uses of an AI system may require its cooperation over multiple turns~\citep{li2024llmdefensesrobustmultiturn, Haize, nakash2024breakingreactagentsfootinthedoor}, and many harmful requests do not produce immediate catastrophic outcomes.
This means that there are more affordances available on the system level to mitigate the vast majority of harms:
\begin{itemize}
    \item \textbf{Trajectory monitoring:} Instead of flagging harmful requests and inducing refusal, several recent approaches apply classifiers to model outputs to filter their responses \cite{inan2023llamaguardllmbasedinputoutput, sharma2025constitutionalclassifiersdefendinguniversal, zeng2024shieldgemmagenerativeaicontent}. However, for harms that accrue over long-term interactions, harm should be classified on entire output \textit{trajectories}, including the history of agent actions and user interactions. Red teaming research on trajectory-level harm is challenging because crafting extended user trajectories that can corner the model, soften its defenses~\citep{russinovich2024great}, or accrue harm over multiple model responses takes time, iterations, and potentially large amounts of resources. As the space of user trajectories for agentic systems grows increasingly large, effectively exploring this attack surface via red teaming is a growing imperative.
    \item \textbf{User monitoring:} Beyond trajectory monitoring, we may monitor \textit{users} of AI systems for patterns of exploitative behavior, including successful inducement of harmful actions detected through post-hoc analysis or real-time observations of repetitive harmful requests. If a small proportion of users represent a large portion of harmful requests, detecting and banning these users can mitigate a great deal of total harm, which can be done using asynchronous review (e.g., by humans, computationally expensive automated systems, or a combination of both). For harms that are diffuse across many requests, such measures may suffice to establish an acceptable level of safety for many cases. For these reasons, safeguard research should invest in malicious \textit{user} detection as well as system-level measures like sybil detection and robust methods for proving user identity.
\end{itemize}

\begin{figure}[t]
    \centering
    \includegraphics[width=0.9\linewidth]{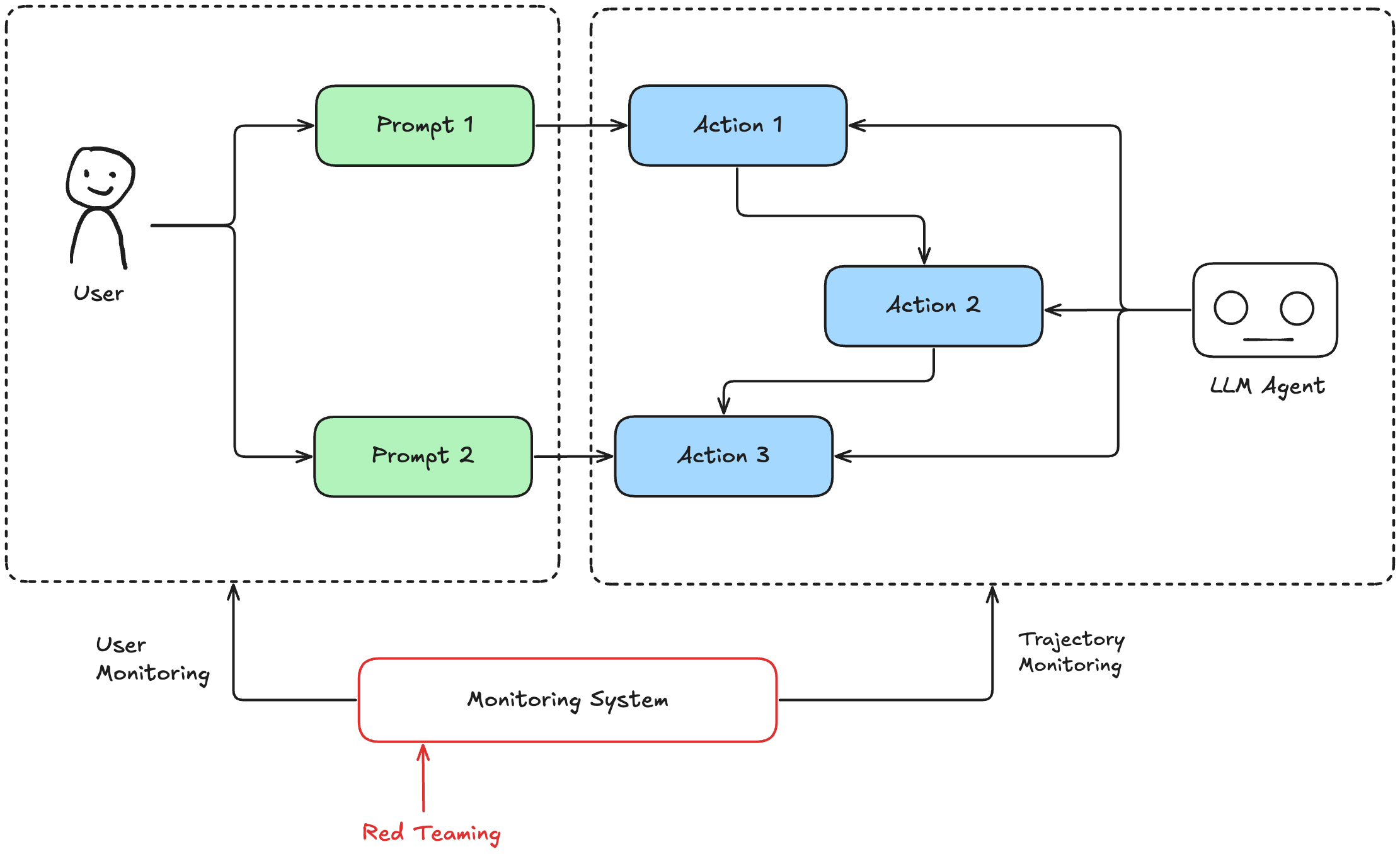}
    \caption{An illustration of user monitoring, trajectory monitoring and monitor red teaming.}
    \label{fig:monitor}
\end{figure}

An illustration is shown in Figure~\ref{fig:monitor}.

\paragraph{Rapid Response to Safeguard Failures.}
AI products built on closed-weight models--and, in some cases, open-weight models as well--do not need to be deployed as static artifacts: as vulnerabilities are detected, they can be patched with new safeguards to prevent future misuse. As in the case of trajectory and user monitoring where harm is diffuse over many queries or single harmful responses are not catastrophic, this fact allows model developers to mitigate the vast majority of harmful outcomes even if model safeguards fail, as long as sufficiently effective asynchronous monitoring detects these failures as they happen. In these cases, rapid response techniques can be employed to quickly develop and deploy safeguards to newly discovered failures before they are exploited at scale~\citep{peng-etal-2024-rapid}. This justifies further research on asynchronous monitoring (e.g., making efficient use of limited human oversight to catch the worst safeguard failures) together with lightweight safeguard adaptation techniques that can be quickly deployed.

\paragraph{Red Teaming The Monitor.} While iterating on the precision and recall of the monitoring system and the effectiveness of intervention strategies is outside the scope of red teaming (and is often considered the "blue team"'s goal), the red team must stress test the monitoring system to help identify potential failures. First, red teams should attempt to complete an adversarial task without being detected by the monitor or any violative behaviors (also referred to as a sabotage experiment)~\citep{benton2024sabotageevaluationsfrontiermodels}. Through red teaming the monitor, we assess the boundary of its effectiveness. Initial research directions should cover the limitation and failure cases for existing simple monitoring systems (e.g. only monitoring the CoT at each step in a trajectory~\citep{baker2025monitoringreasoningmodelsmisbehavior}) with well-crafted adversarial techniques, as well as suggestions to improve them that feed into the blue team's work. 

\section{Alternative Views}

\paragraph{Argument for Static Harm Research.} 

This framework takes a broad view of AI safety, acknowledging that some released models will simply have a different policy and be much more likely to engage in harm based on the developer’s jurisdiction, philosophy, and intent. However, while a policy-based evaluation, coupled with widespread vulnerability probing to detect failures the policy potentially overlooks, makes sense for a robust and replicable understanding of AI behavior, it may not fully account for understanding the potential risks that AI may pose to society writ large. To assess these larger questions, researchers should instead deliberately adopt a normative framework for harm definitions, perhaps guided to a clear and aligned industry standard, and document the ways in which AI may violate these standards, both with respect to direct requests and other manipulation methods. Open-weight models are also more difficult to red team since fine-tuning can effectively modify their behaviors, such as removing guardrails from model developers. Regulation may change the landscape, but such products are likely to always exist in some form.

The challenge with this framework is that it separates from replicable model behaviors and requires researchers to engage in normative definitions. The ethics of what content is harmful can be extremely context-dependent, sensitive, and lack consensus. Legal definitions, researcher assumptions, and commonly held beliefs may not hold up. 

\section{Conclusion}
In this position paper, we argue that red teaming should prioritize product safety specifications with realistic threat models over abstract social biases, and system-level safety over model-level robustness. Because model developers and downstream products deployers can always have distinct safety specifications, red teaming should align the with intended goal of the product instead of abstract and general harm on the model-level. We further present a set of examples of realistic threat models for technical researchers to study and improve the attack techniques and a set of good practices in doing red teaming. Finally, we argue that safeguard  research should expand to scope to the system level, monitoring and ensuring harmless interactions between the AI and human users.

\section*{Acknowledgment}

We appreciate the feedback from Niall Dalton, Nathaniel Li, Felix Binder and Mohamed Shaaban on the early draft of this paper.

\bibliography{custom}
\bibliographystyle{abbrvnat}

\appendix

\end{document}